\documentclass{iopjournal}

\usepackage[utf8]{inputenc}
\usepackage{amsmath}
\usepackage{lmodern}
\usepackage{amssymb}
\usepackage{cite}

\newcommand{\eg}{e.g., }
\newcommand{\ie}{i.e., }

\begin{document}

\articletype{Paper}

\title{Particle dynamics and confinement in moving multi-mirror}

\author{Tal Miller$^1$\orcid{0000-0002-3636-7141},  Eli Gudinetsky$^1$\orcid{0009-0002-4310-9818}, Ilan Be'ery$^2$\orcid{0000-0003-3405-9898}, and Ido Barth$^{1,*}$\orcid{0000-0001-8198-896X}}

\affil{$^1$Racah Institute of Physics, The Hebrew University of Jerusalem, Jerusalem, 91904 Israel}

\affil{$^2$nT-Tao, 5 Ha-Nagar st., Ramat Hasharon, 4526005 Israel }

\affil{$^*$Author to whom any correspondence should be addressed.}

\email{ido.barth@mail.huji.ac.il}


\begin{abstract}

The moving multi-mirror (MMM) concept mitigates axial losses in magnetic mirrors by using inward-propagating multi-mirror sections to transport escaping particles back toward the central cell.
Single-particle simulations are used to establish the underlying dynamics, while a modified rate-equation model provides quantitative estimates of the resulting confinement enhancement.
It is found that the steady-state outgoing flux can be robustly suppressed by several orders of magnitude.
The analysis uncovered confinement challenges in the MMM concept, indicating that additional scattering processes are required in both the central cell and the MMM sections to achieve the desired confinement.

\end{abstract}
\section{Introduction}
\label{sec: intro}

Magnetic mirror machines offer several attractive features as potential fusion reactors, including steady-state operation, high-$\beta$, and relatively slow cross-field transport \cite{post1987magnetic, ivanov2003experimental, burdakov2010modern, beklemishev2010vortex, be2015feedback}.
However, their performance is limited by particle and energy losses through the mirror's loss cone. 
In the past $60$ years, various static and dynamic plugging methods have been introduced, including tandem mirror machines with thermal barriers \cite{inutake1985thermal, grubb1984thermal, katanuma1986thermal, pratt2006global, tamano1995tandem, ivanov2013gas, ivanov2017gas, egedal2022fusion, endrizzi2023}, radio-frequency (RF) plugs \cite{golovato1985fueling}, diamagnetic confinement \cite{beklemishev2016diamagnetic, kotelnikov2020structure}, 
ponderomotive RF plugging \cite{motz1967radio, watari1974theory, hatori1975critical, watari1978radio, uehara1978radio, hiroe1978experiment, fader1981rf, fisch2003current, dodin2004ponderomotive, dodin2005ponderomotive}, and side-sections of field-reversed configuration (FRC) at the mirror throats \cite{shi2019magnetic}.
Additionally, static MM sections were proposed decades ago to convert these losses into a diffusive process, thereby lengthening the effective confinement time \cite{post1967confinement, logan1972multiple, logan1972experimental, mirnov1972gas, makhijani1974plasma, tuszewski1977transient, burdakov2016multiple, budker1971influence, mirnov1996multiple, kotelnikov2007new,post1981particle}. 
Still, the confinement achievable with static MM is insufficient for fusion conditions, \ie the Lawson criterion \cite{post1967confinement, kotelnikov2007new, lawson1957some, miller2021rate, wurzel2022progress}. 

\begin{figure}
    \centering
    \includegraphics[clip, trim=0.0cm 0.0cm 0.0cm 0.0cm, width=0.5\linewidth]{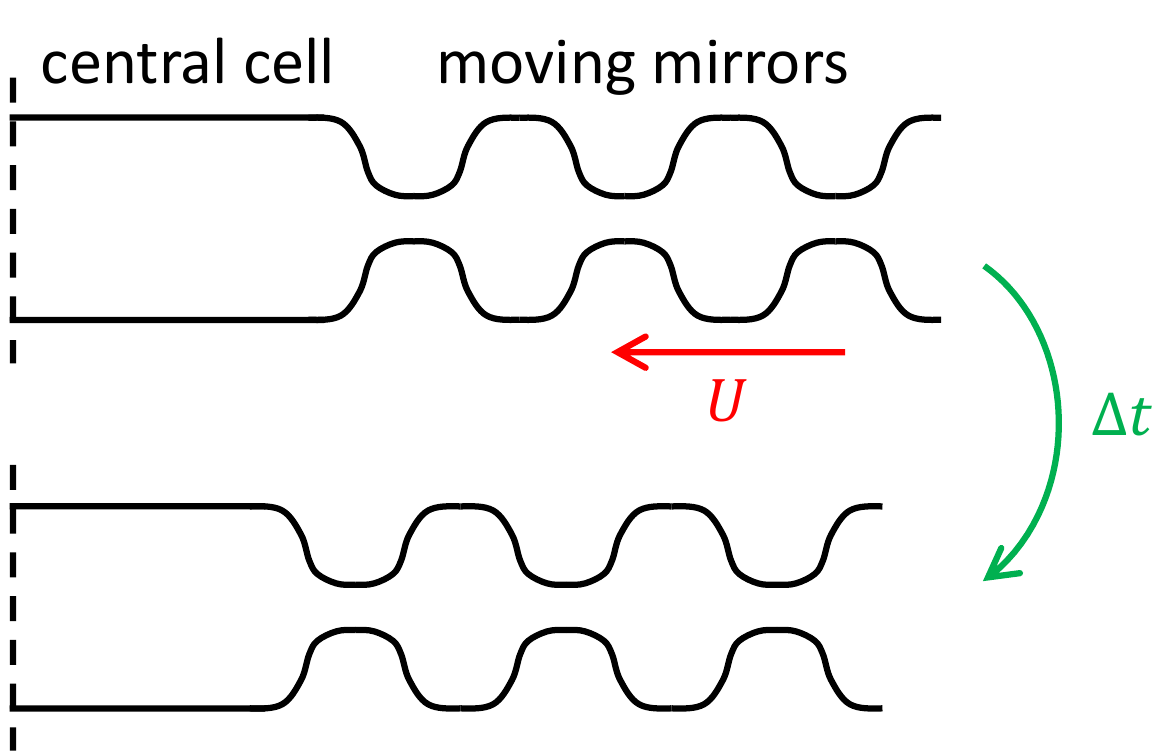}
    \caption{An illustration of an MMM system. The upper and lower panels show the schematic magnetic field lines at two different times as the MMM section moves.}
    \label{fig: MMM schematic}
\end{figure}

One interesting direction, first introduced in \cite{tuck1968reduction, budker1982gas}, is the use of a moving multi-mirror (MMM) section. 
In this concept, the MM field structure is electromagnetically driven toward the main cell, creating a pumping mechanism that opposes axial losses by transporting confined particles from the MM sections back toward the central cell, thereby increasing overall confinement. 
A schematic illustration of the MMM concept is presented in Fig~\ref{fig: MMM schematic}.
A similar method is the helical mirror, where a stationary helically symmetric magnetic field acts as a moving mirror system in the reference frame of a rotating plasma, resulting in an axial pumping force \cite{beklemishev2013helicoidal, postupaev2016helical, sudnikov2019first, ivanov2021long, sudnikov2022plasma,tolkachev2024electromagnetic, sudnikov2024improved}.

Recently, the MMM concept was revisited in \cite{be2018plasma}, presenting a quantitative engineering analysis. 
By driving a propagating magnetic-wave structure along an MM chain, they analytically showed that the axial flux could be suppressed well below the level required by the Lawson criterion, with voltage and current values readily accessible by modern solid-state switches. 
They also outlined a compact laboratory demonstrator capable of demonstrating a two-order-of-magnitude reduction in end losses. 
However, the analysis relied on a fluid-like diffusion approximation and did not address several microscopic issues critical to realistic implementation.
These include the detailed electromagnetic fields required to realize an MMM section, the particle dynamics near the interface between the MMM section and the central cell, the modification of loss-cone boundaries in the moving reference frame, and a kinetic description of the population dynamics.
 
The present paper addresses these issues and thereby advances the microscopic and kinetic foundations of the MMM concept. 
To this end, we construct an explicit, Maxwell-consistent field realization of a full mirror-MMM system and simulate single-particle trajectories to reveal how particles are swept, accelerated, or lost at the interface with the central cell.
We also introduce a laboratory-frame, modified rate-equation model by incorporating the velocity-dependent loss-cone solid angles and an effective MMM drag term into our previous MM framework, and solve the resulting coupled equations for the steady-state particle flux under different assumptions regarding the plasma velocity distribution.
We find that the resulting confinement enhancement is robust, reaching several orders of magnitude for realistic system parameters, 
provided an additional effective scattering mechanism is available for the escaping and the recaptured particles.

The structure of the paper is as follows.
Sec.~\ref{sec: single particle} proposes a setup for the magnetic field and the induced electric field for a system composed of a central fusion cell and two MMM sections.
In this section, we also perform single-particle calculations to illustrate representative trajectories, examine the stitching problem, and propose a first step toward its possible solution.
In Sec.~\ref{sec: loss cone}, we analyze the effective loss cone of a moving mirror in a static reference frame.
Sec.~\ref{sec: rate equations} incorporates the effect of MMM-induced transport into a modified rate-equation model for the MM system and calculates the confinement enhancement, while Sec.~\ref{conclusions} discusses and summarizes the findings.

\section{Field setup and single-particle trajectories}
\label{sec: single particle}

\begin{figure}
    \centering
    \includegraphics[clip, trim=0.0cm 0.0cm 0.0cm 0.0cm, width=\linewidth]{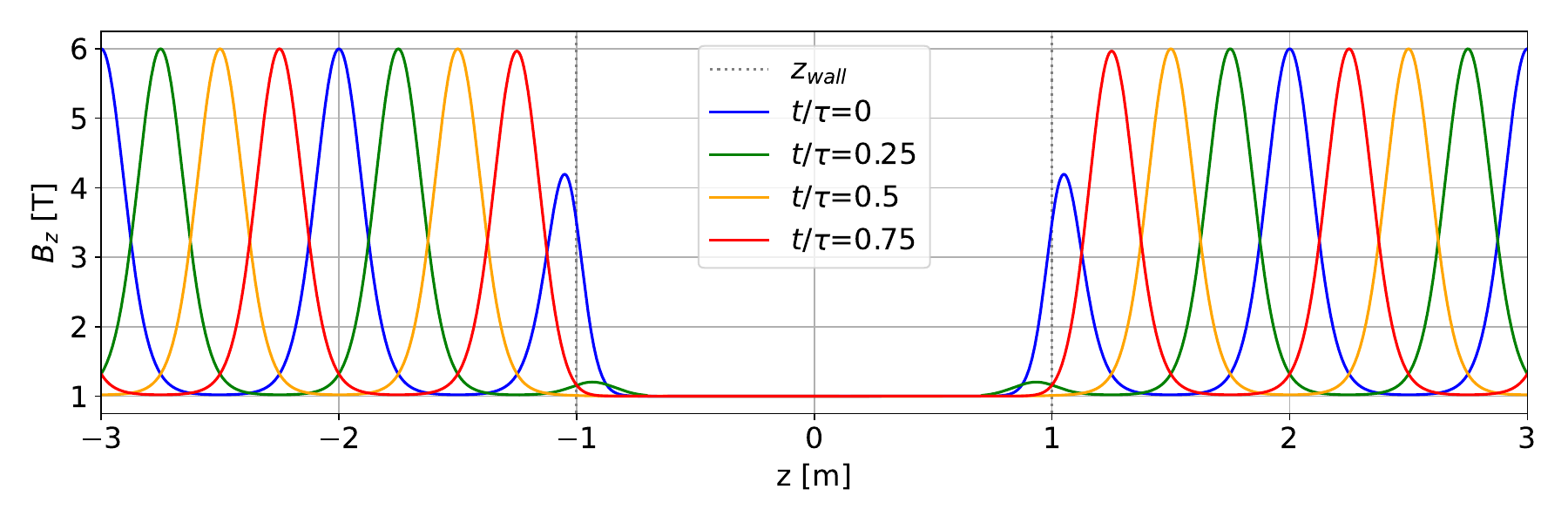}
    \includegraphics[clip, trim=0.0cm 0.0cm 0.0cm 0.0cm, width=\linewidth]{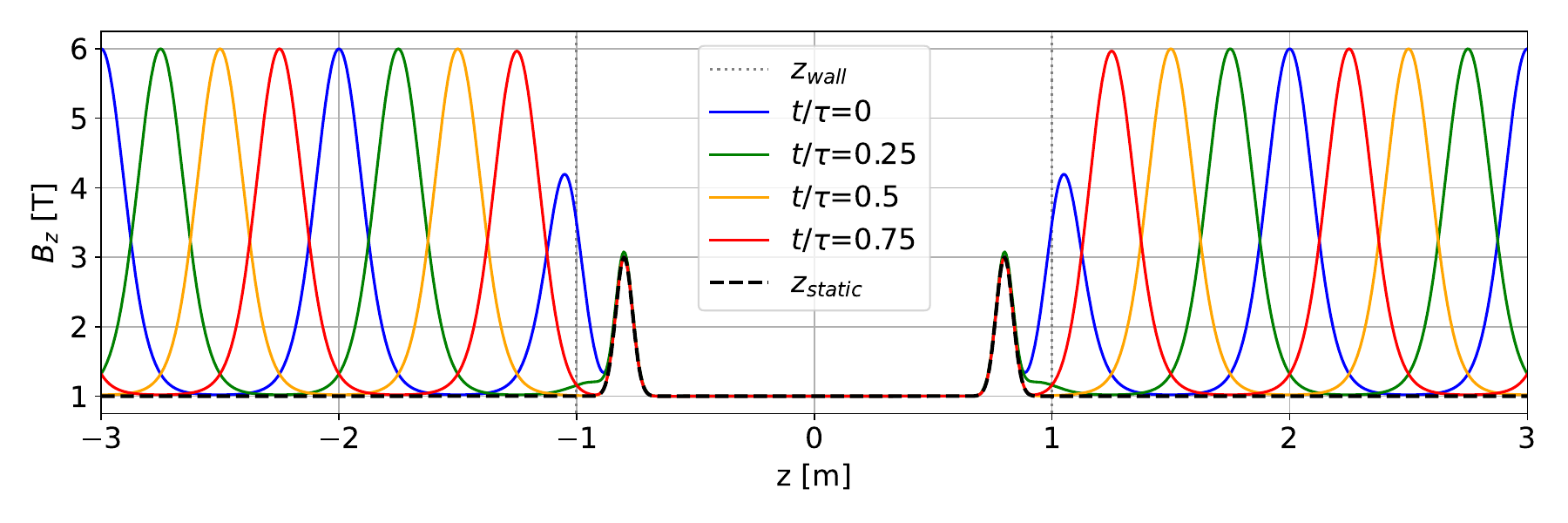}
    \caption{Axial magnetic field in the full MMM configuration comprising the central cell and moving sections for a mirror ratio $R_m=6$, without (top) and with (bottom) an additional static central cell field (with $\widetilde R_m=3$). 
    The colors indicate the dynamic MMM field profile at different times (see legend).}
    \label{fig: full MMM axial B field}
\end{figure}

The plugging mechanism of an MMM section is based on the inward transport of trapped particles in each MM cell induced by the propagation of the mirror structure. 
A key unresolved question, however, concerns the interface between the MMM section and the central cell.
Specifically, how the magnetic fields of these two regions are "stitched" together, and what is the dynamics of trapped particles in the central cell when experiencing the time-dependent magnetic field at the interface.
To the best of our knowledge, this problem has not been addressed in the literature. 
To investigate this, we construct a self-consistent, simplified model for the magnetic and induced electric fields in both the central cell and the MMM section, and analyze the resulting dynamics using single-particle simulations.

Let us begin with a realization of a static (MM) magnetic field suggested in \cite{post1967confinement}.
The axial ($\hat{z}$) component reads 
\begin{eqnarray}
    B_\text{MM} \left(z \right)=B_{0}\left[1+\left(R_{m}-1\right)\exp\left(-5.5\sin^{2}\frac{\pi z}{l}\right)\right]
    \label{eq: Bz mirror}
\end{eqnarray}
where $R_m$ is the mirror ratio and $l$ is the length of one MM cell. 
The radial component is determined by Gauss's law of magnetism, $\nabla \cdot B=0$.
It is notable that this field is time-independent and, therefore, does not induce an electric field.

To model the MMM axial field as a moving MM field, we replace the argument in $B_\text{MM}(z)$ with a moving coordinate of velocity, $U$.
So, for example, the axial component of the magnetic field in the right MMM section reads $B_\text{MM}(z+Ut)$ as illustrated in Fig.~\ref{fig: MMM schematic}.
To ensure a smooth interface with the central cell, we introduce a cutoff function $W\left( z \right)$ that suppresses the time-dependent fields within the central cell region, such that the magnetic field smoothly saturates to the minimal value of the central mirror.
The resulting axial component of the magnetic field in the right half of the system can then be written as

\begin{equation} \label{B_z}
    B_z \left(z,t \right) = B_0 + \left(B_{MM} \left(z+Ut \right)-B_0\right) W\left( z \right)
\end{equation}
The corresponding radial component of the field, satisfying Gauss’s law, is therefore

\begin{equation}\label{B_r}
    \begin{split}
        B_r (z,r,t) &= -\frac{r}{2} \frac{\partial B_z}{\partial z} \\
        &= -\frac{r}{2} \left[ \frac{\partial B_{MM}(z+Ut)}{\partial z} W(z) 
            + \left(B_{MM}(z+Ut) - B_0\right) \frac{\partial W(z)}{\partial z} \right],
    \end{split}
\end{equation}
and the induced electric field, obtained from Faraday's law, is
\begin{equation} \label{E_theta}
    E_{\theta} \left(z,r,t \right)=-\frac{r}{2} U \frac{\partial B_{MM}\left(z+Ut \right)}{\partial z} W\left(z \right).
\end{equation}
Finally, we define the interface smooth suppressing function as
\begin{equation} \label{W_z}   
    W\left( z \right) =\left[ 1 + \exp \left( -\frac{z-z_\text{wall}}{\Delta z_\text{wall}} \right) \right]^{-1},
\end{equation}
where $z_\text{wall}$ and $\Delta z_\text{wall}$ are the axial location and width of the interface between the central fusion cell and the MMM section, respectively. 
By superposing two MMM sections with parameters $U,z_\text{wall}$ and $-U,-z_\text{wall}$, we obtain a symmetric MMM system with the central cell located at $z=0$ and MMM pumping sections on both sides.

\begin{figure}
    \centering
    \includegraphics[clip, trim=0.0cm 0.0cm 0.0cm 0.0cm, width=\linewidth]{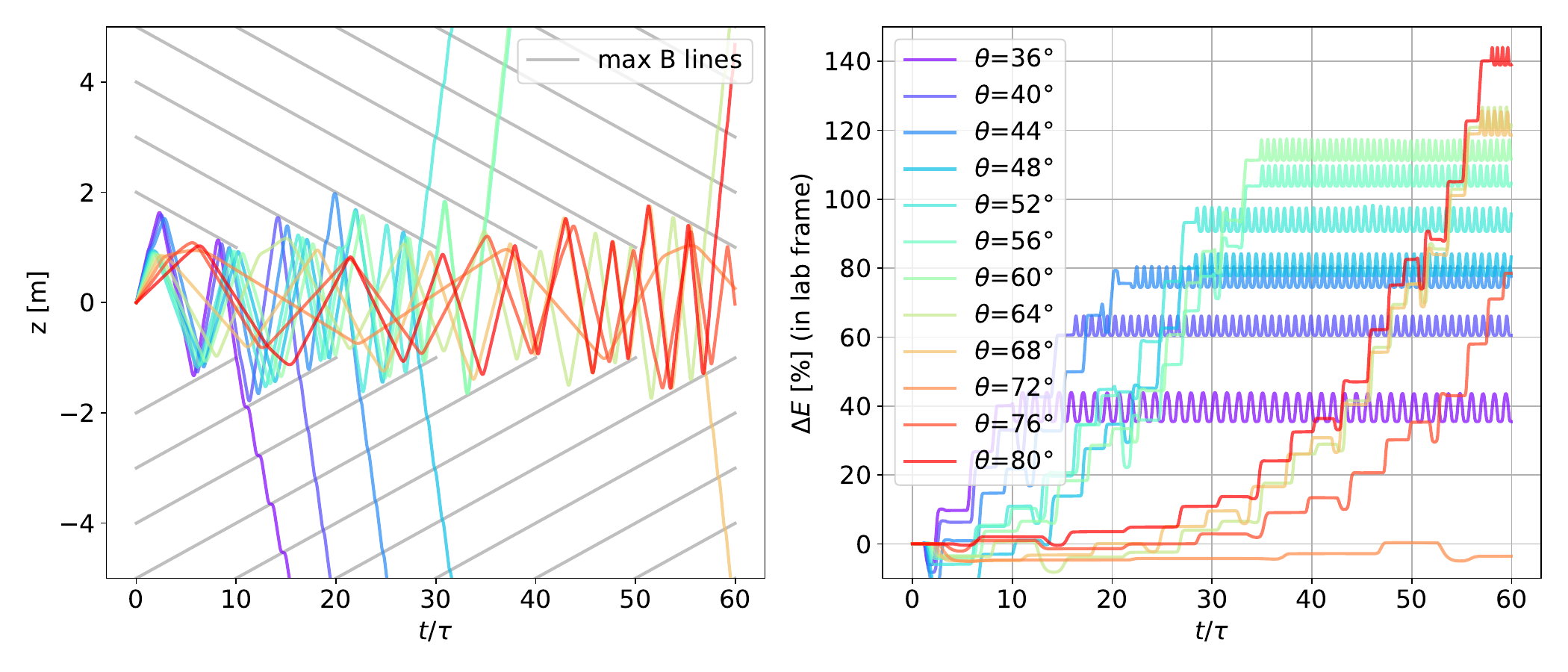}
    \hfill
    \centering
    \includegraphics[clip, trim=0.0cm 0.0cm 0.0cm 0.0cm, width=\linewidth]{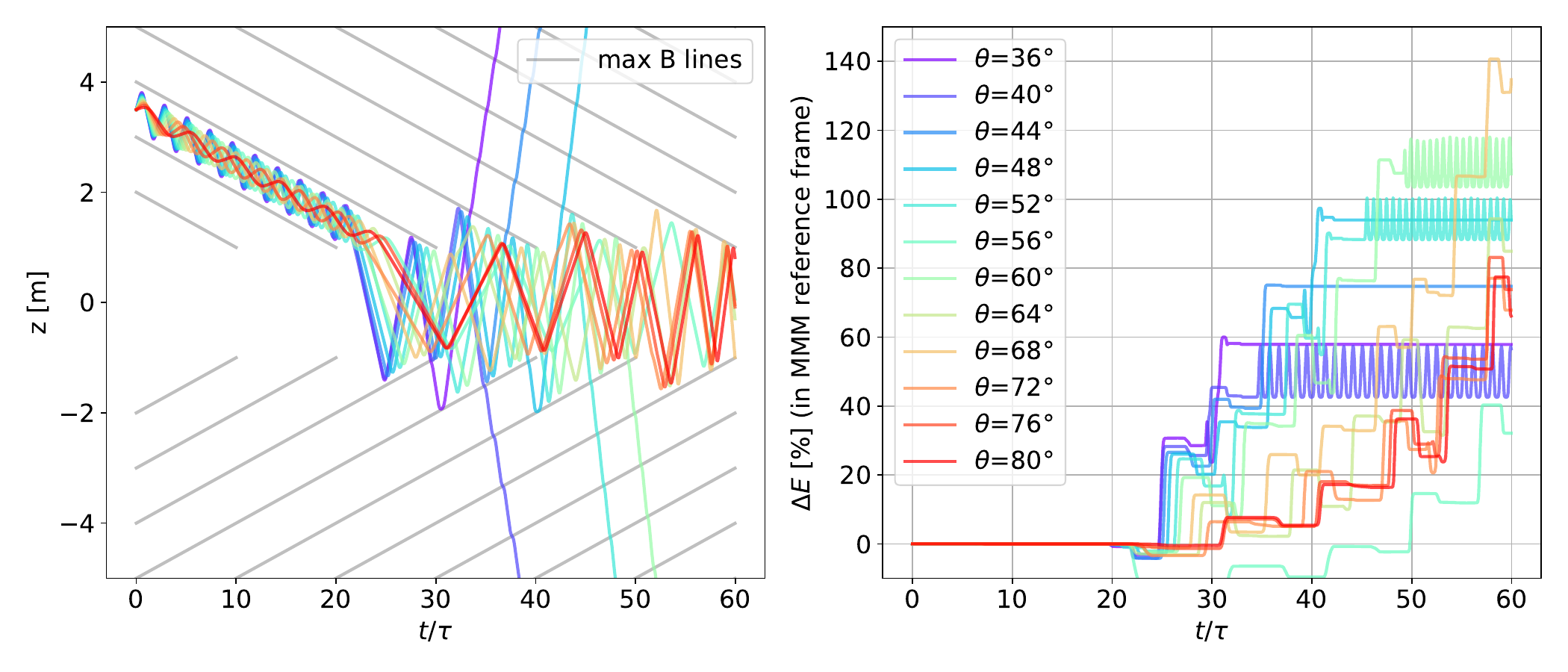}
    \caption{Single-particle trajectories initialized as trapped in the center of the central cell (top) and in the MMM section (bottom). 
    Left: time evolution of the axial particle positions (colored curves) relative to the locations of the magnetic field maxima (gray lines). 
    Right: corresponding particle energy evolution.
    The parameters used are $R_m=6$, $z_\text{wall}=1$~m, $\Delta z_\text{wall}=0.05$~m, $l=1$~m, $U=0.1v_{th}$.}
    \label{fig: single particle trajectories}
\end{figure}

\begin{figure}
    \centering
    \includegraphics[clip, trim=0.0cm 0.0cm 0.0cm 0.0cm, width=\linewidth]{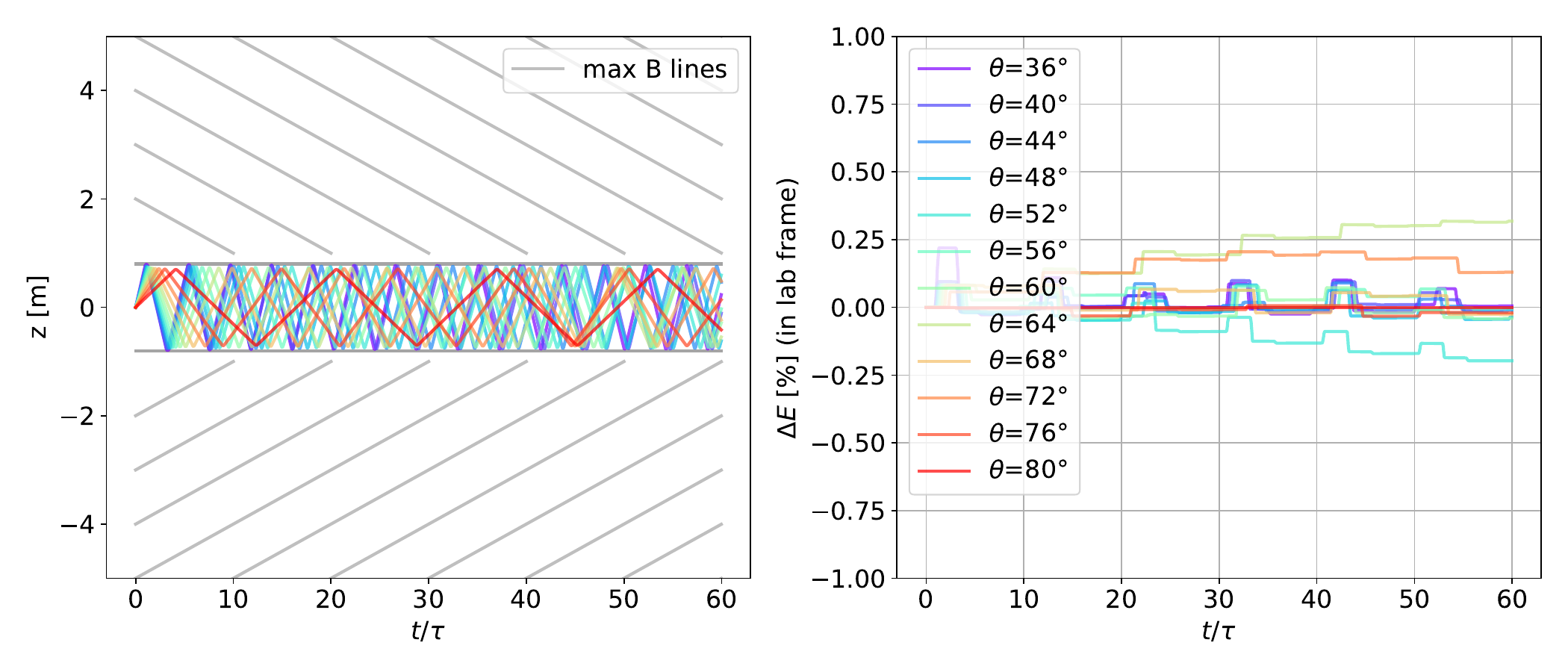}
    \hfill
    \centering
    \includegraphics[clip, trim=0.0cm 0.0cm 0.0cm 0.0cm, width=\linewidth]{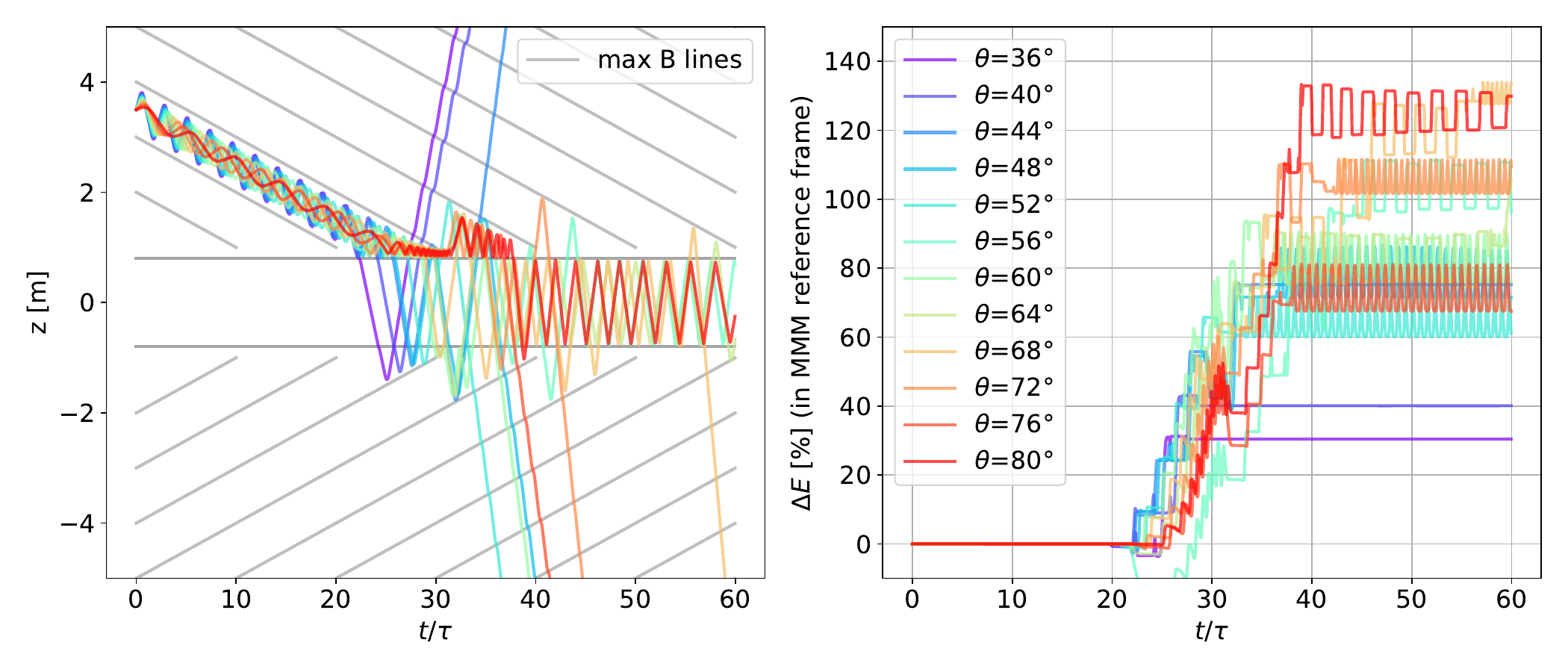}

    \caption{Single-particle trajectories initialized as trapped in the center of the central cell (top) and in the MMM section (bottom) for the modified magnetic field that includes an additional static central cell field.
    Left: time evolution of the axial particle positions (colored curves) relative to the locations of the dynamic and static magnetic field maxima (gray lines). 
    Right: corresponding particle energy evolution.
    The parameters used are $R_m=6$, $z_\text{wall}=1$~m, $\Delta z_\text{wall}=0.05$~m, $l=1$~m, $U=0.1v_{th}$, and for the static main cell $\widetilde R_m=3$, $z_\text{static}=0.8$~m.}
    \label{fig: single particle trajectories with static main cell}
\end{figure}

For illustration, we consider an MMM system with the following representative parameters: $l=1$~m,  $R_m=6$, and $B_0=1$~T.
The wall of the static central cell is located at $z_\text{wall}=1$~m, with width $\Delta z_\text{wall}=0.05$~m.
Although fusion-relevant systems would typically employ a much longer central cell, a shorter length is used here to reduce computational cost, without affecting the qualitative conclusions regarding the MMM plugging effect.
The MMM propagation velocity is $U=0.1v_{th}=1.37\cdot 10^5$~m/s, where $v_{th}$ is the thermal velocity hydrogen (protium) at $T=10$~keV.
In Fig.~\ref{fig: full MMM axial B field} (upper panel), we plot the full axial component of the magnetic field, $B_z$, at $r=0$ for four different time frames defined by the normalizing timescale, $\tau=l/v_{th}$.
It can be seen that both MMM sections propagate inward toward the central cell, analogous to a pump or a conveyor belt.

Next, we perform single-particle simulations to study the dynamics of the fuel ions in the MMM system for two classes of particles, one trapped in the central (fusion) cell and the other in one of the MMM cells.
For the first class of particles, we select $11$ initial velocity directions characterized by the pitch angle $\theta$ relative to the mirror axis, in the range $\left[36 ^\circ,80 ^\circ \right]$.
This range lies outside the loss cone of a static mirror with $R_m=6$ for which the loss-cone angle is
$\theta \approx 24.1 ^\circ$.
The effect of trapping by moving magnetic walls is discussed below. 
All particles are initialized at the center of the central cell, $z=0$, and with identical speed, $v_{th}$. 
For the second class of particles, we use the same initial conditions but initialize them at the center of one of the MMM cells, at $z=3.5$~m, and shift their axial velocities according to $v_z \rightarrow v_z - U$.
These particles are, by construction, trapped in the MMM reference frame.
We numerically simulate the trajectories of particles from both classes under the influence of the time-dependent MMM fields defined in Eqs.~(\ref{B_z}-\ref{E_theta}) over a time interval of $60\,\tau$. 

The simulation results are presented in Fig.~\ref{fig: single particle trajectories} for both the central cell particles (upper panels) and the MMM cell particles (lower panels).
The results indicate that particles initially confined in the central cell, \ie of the first class, do not follow simple static-mirror dynamics, as the mirror effectively expands and contracts. 
Instead, the particles interact with a counter-propagating moving mirror and are accelerated along the axial direction at each reflection, resulting in a systematic energy gain, as observed in the energy evolution plots (upper right panel of Fig.~\ref{fig: single particle trajectories}). 
This behavior is analogous to the classical (second-order) Fermi acceleration mechanism.
Given sufficient time, the particles accumulate axial energy until they eventually escape through one of the loss cones.
At the same time, particles of the second class, \ie originating in the MMM cell, are swept inward as expected, conserving their energy in the MMM reference frame.
However, upon entering the central cell, they undergo the same acceleration mechanism, causing their axial energy to continue increasing until they eventually escape.

To mitigate this effect, we introduce an additional static magnetic field at the edges of the central cell, whose axial component reads
\begin{equation} \label{B_static}   
    B_\text{static}\left(z \right)= \sum_{\eta=\pm 1} B_0 \left( \widetilde R_{m} -1 \right) \exp \left( -\left( \frac{z-\eta\, z_\text{static}}{\Delta z_\text{wall}} \right)^2 \right),
\end{equation}
where $ \widetilde{R}_m =3$ is the mirror ratio of the static barrier, smaller than that of the MMM section. 
The field terminates at $z_\text{static}=0.8$~m, slightly shorter than $z_\text{wall}=1$~m. 
We also use the same parameter $\Delta z_\text{wall}$ from Eq.~(\ref{W_z}).
As in Eq.~(\ref{B_r}), Gauss's law of magnetism determines the corresponding radial component, $B_r = -\frac{r}{2} \frac{\partial B_\text{static}}{\partial z}$.
The discrete parameter, $\eta=\pm1$, labels the left and right components of the static barrier.
The bottom panel of Fig.~\ref{fig: full MMM axial B field} illustrates the additional static magnetic field (dashed black line) and the total axial field at different times (colored lines).

We repeat the single-particle simulations for both classes of particles using the same initial conditions as in Fig.~\ref{fig: single particle trajectories}, but with the modified field that includes the static field of Eq.~(\ref{B_static}).
Note that the particles of the first class, \ie central cell particles, would be trapped in the static mirror with a mirror ratio of $\widetilde R_m=3$ since the corresponding loss-cone angle is $\theta \approx 35.3 ^\circ$ and all simulated particles have larger initial pitch angles.
The results are presented in Fig.~\ref{fig: single particle trajectories with static main cell}, showing that the initially trapped particles experience standard mirror dynamics (upper left panel) with a little energy change (upper right panel). 
At the same time, the particles in the MMM section are swept inward (bottom left panel) but experience a significant increase in energy upon merging into the main cell (bottom right panel) due to pinching between magnetic-field barriers on both sides.
As a result, their axial energy exceeds the static mirror barrier, so they are confined only by the dynamic MMM fields,
thereby remaining subject to axial acceleration and reduced confinement.
This severe limitation will be discussed in Sec.~\ref{conclusions}, but before doing so, we analyze the effective loss cone for the MMM system in the next section and develop a modified rate-equation model in Sec.~\ref{sec: rate equations}.

\section{Loss-cones in a moving mirror}
\label{sec: loss cone}

\begin{figure}
    \centering
    \includegraphics[clip, trim=0.0cm 0.0cm 0.0cm 0.0cm, width=0.8\linewidth]{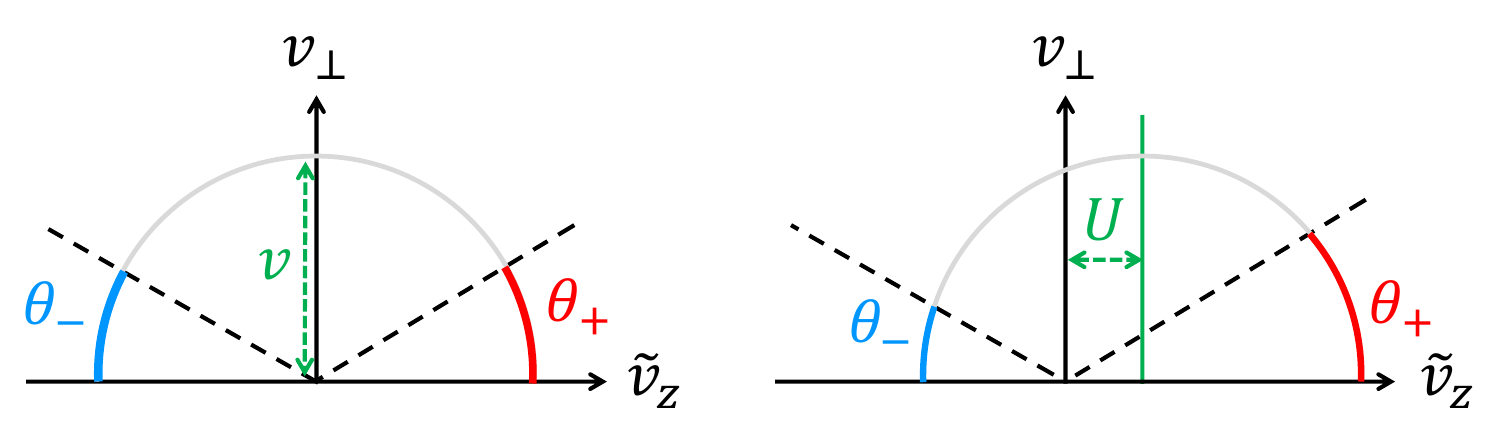}
    \includegraphics[clip, trim=0.0cm 0.0cm 0.0cm 0.0cm, width=0.8\linewidth]{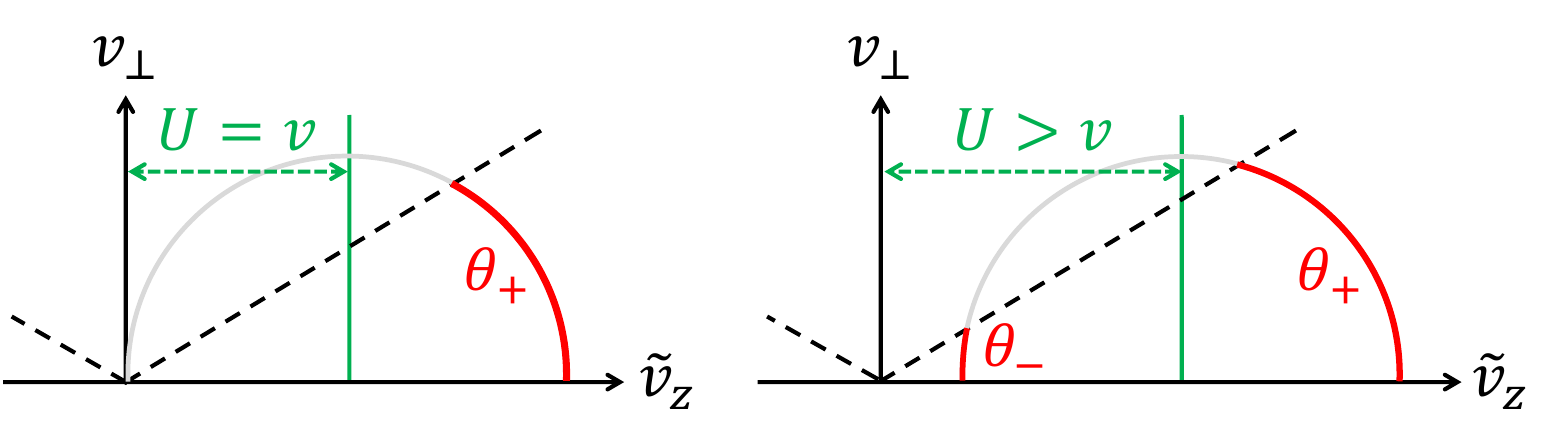}
    \caption{Schematic illustration of the loss-cone solid angles in the laboratory frame. 
    The colored semicircle is the equi-velocity curve in the (effective) laboratory frame, illustrating the three populations (see Eqs.~(\ref{alpha_l})-(\ref{alpha_c})): captured (gray), right-going (red), and left-going (blue) loss cones.
    The dashed lines indicate the loss cones that are defined (symmetrically) in the moving frame, $\tilde{v}_z \equiv v_z + U$, where $U$ is the inward (left-going) mirror speed.
    The different panels illustrate four different values of the mirror velocity, $U$:
    static mirror, $U=0$ (top-left panel); $U<v$ (top-right); $U=v$ (bottom-left); and $v<U<U_m$ (bottom-right).}
    \label{fig: MMM loss cone schematic}
\end{figure}

When a magnetic mirror propagates axially, the classification of particles as trapped or belonging to one of the loss cones must be evaluated in the mirror reference frame, provided that the magnetic moment is adiabatically conserved.
For example, for a mirror propagating to the left, the loss-cone boundaries become asymmetric in the laboratory frame, where the left loss cone shrinks, while the right loss cone expands.
Consequently, when the MM structure moves toward the main cell, intended to pump particles inward, this asymmetry is detrimental, as it enlarges the outward loss cone and enhances particle escape.
This effect is illustrated schematically in Fig.~\ref{fig: MMM loss cone schematic}. 
We next quantify this effect by calculating the solid angles of the modified loss cones as a function of the mirror velocity, $U$.

Recall that the loss cone condition for a particle with transverse and axial velocities $v_{\perp},v_{z}$ in the midplane of a static mirror is $v_{\perp}^{2}/v^{2}<R_{m}^{-1}$, where  $v^{2}=v_{\perp}^{2}+v_{z}^{2}$. 
For an MMM propagating to the left, we define the MMM axial velocity,  $\mathbf{v}_\text{\tiny{MMM}}=-U\hat{z}$, where $U>0$.
Thus, in the moving mirror frame, the particle axial velocity becomes $\tilde{v}_{z}=v_{z}+U$, while the total velocity square of the moving frame is $\tilde{v}^2=v_{\perp}^{2}+\tilde{v}_{z}^{2}$ as the transverse velocity is invariant to this transformation.
In equilibrium, the mirror escaping condition is determined in the rest frame of the moving mirror, so the loss-cone condition in the mirror frame reads
\begin{equation}
\label{eq: loss cone condition moving}
\frac{v_{\perp}^2}{\tilde{v}^2} <\frac{1}{R_{m}}.
\end{equation}
 
The effective rate-equation model for MMM systems, to be introduced in the next section, is formulated in the static laboratory frame and incorporates the mirror dynamics only through effective transport terms.
Therefore, we evaluate the loss-cone condition here in terms of laboratory frame velocities.
Using the laboratory frame velocity variables, Eq.~(\ref{eq: loss cone condition moving}) becomes
\begin{equation}
\label{eq: loss cone condition lab}
\frac{v_{\perp}^{2}}{v_{\perp}^{2}+\left(v_{z}+U\right)^{2}} < \frac{1}{R_{m}}.
\end{equation}
To extract the solid angle we solve it for $v_{z}$ using the identity $v_{\perp}^{2}=v^{2}-v_{z}^{2}$
\begin{equation}
R_{m}\left(v^{2}-v_{z}^{2}\right) = v^{2}-v_{z}^{2}+\left(v_{z}+U\right)^{2}.
\end{equation}
Solving the quadratic equation for $v_z$, one finds two solutions,
\begin{equation} \label{eq: v_z}
v_{z,\pm}=-\frac{U}{R_{m}} \pm \frac{\sqrt{ R_{m}-1} }{R_{m}} \sqrt{v^{2}R_{m}-U^{2}}.
\end{equation}
Requiring the discriminant to be non-negative yields the necessary condition $\left|U\right|\leq\sqrt{R_{m}}v$.
We therefore define the corresponding critical velocity,
\begin{equation}
U_\text{last}=\sqrt{R_{m}}v, 
\end{equation}
beyond which no loss-cone solution exists. 
The corresponding modified loss-cone limiting angles are obtained from Eq.~(\ref{eq: v_z}) as
\begin{equation} \label{eq: loss cone angles}
\sin\theta_{\pm}=\frac{v_{\perp}}{v}=\frac{\sqrt{v^{2}-v_{z,\pm}^{2}}}{v}.
\end{equation}

The solid angles (over $4\pi$) of the modified loss cones as a function of $U$ depend on the loss cone limiting angles Eq.~(\ref{eq: loss cone angles}) and have different scenarios, which are illustrated in Fig.~\ref{fig: MMM loss cone schematic}.
For moderate MMM velocities, $U<v$, the modified right and left loss cones are shaped like a cone (top-right panel), where the right-going angle increases while the left-going angle decreases in comparison to the static $(U=0)$ mirror (top left panel).
When $U=v$, the left loss cone shrinks to zero, while the right loss cone continues to grow (bottom-left panel).
For higher velocities, $U>v$, a second branch of the right-going loss cone emerges (bottom-right panel).
This branch corresponds to non-trapped left-going particles, but with axial velocity smaller than the MMM velocity, and thus are left behind.
Assuming $v$ is the thermal velocity of the particle population, such high values of $U$ are generally undesirable, as they reduce the plugging efficiency and may lie beyond practical operating limits for fusion-relevant plasmas.
Nevertheless, this regime may remain relevant for particles in the low-velocity tail of the distribution, requiring a detailed kinetic treatment that lies beyond the scope of the present work.
For completeness, we retain this scenario in our analysis.

Combining the results above, the normalized solid angles of the modified loss-cone regions, denoted by $\alpha_r, \alpha_l, \alpha_c$, are given by
\begin{eqnarray}
\label{alpha_l}
\alpha_{l} & = & \begin{cases}
\sin^{2}\left(\frac{\theta_{-}}{2}\right) & U<v\\
0 & U>v
\end{cases}\\
\label{alpha_r}
\alpha_{r} & = & \begin{cases}
\sin^{2}\left(\frac{\theta_{+}}{2}\right) & U<v\\
\sin^{2}\left(\frac{\theta_{+}}{2}\right) + \sin^{2}\left(\frac{\theta_{-}}{2}\right) & v<U<U_{m}\\
1 - \sin^{2}\left(\frac{\theta_{+}}{2}\right)  + \sin^{2}\left(\frac{\theta_{-}}{2}\right) & U_{m}<U<U_{last}\\
1 & U>U_{last}
\end{cases}\\
\label{alpha_c}
\alpha_{c} & = & 1-\alpha_{r}-\alpha_{l}
\end{eqnarray}

\begin{figure}
    \centering
    \includegraphics[clip, trim=0.0cm 0.0cm 0.0cm 0.0cm, width=0.32\linewidth]{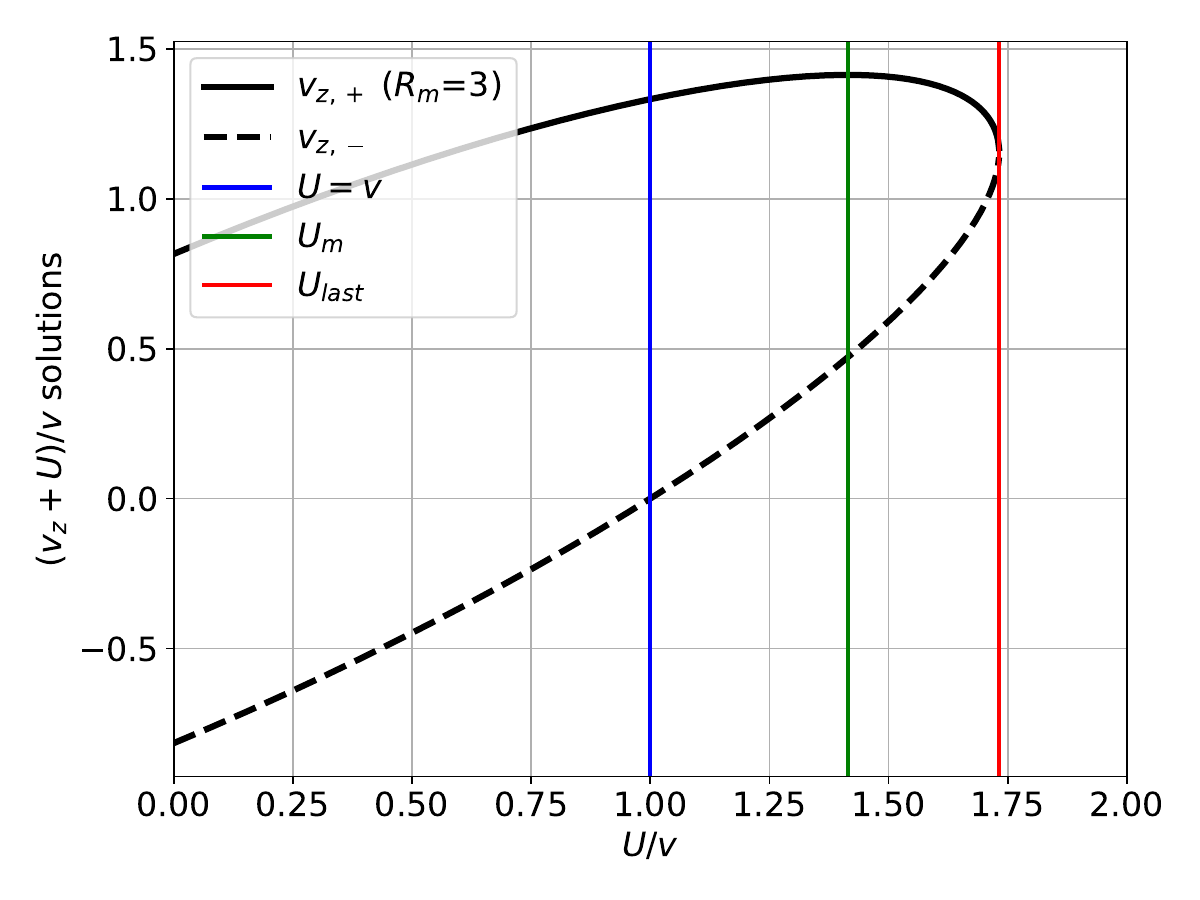}
    \includegraphics[clip, trim=0.0cm 0.0cm 0.0cm 0.0cm, width=0.32\linewidth]{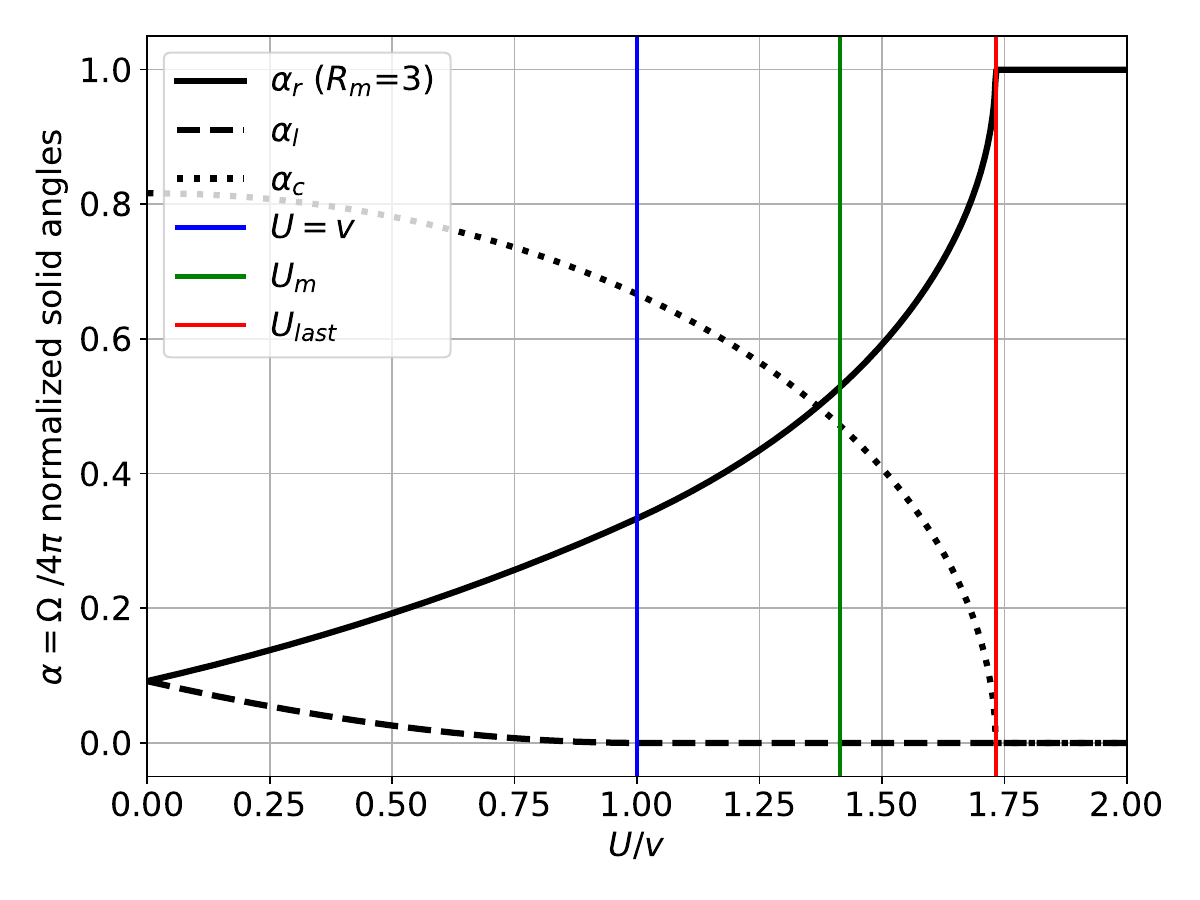}
    \includegraphics[clip, trim=0.0cm 0.0cm 0.0cm 0.0cm, width=0.32\linewidth]{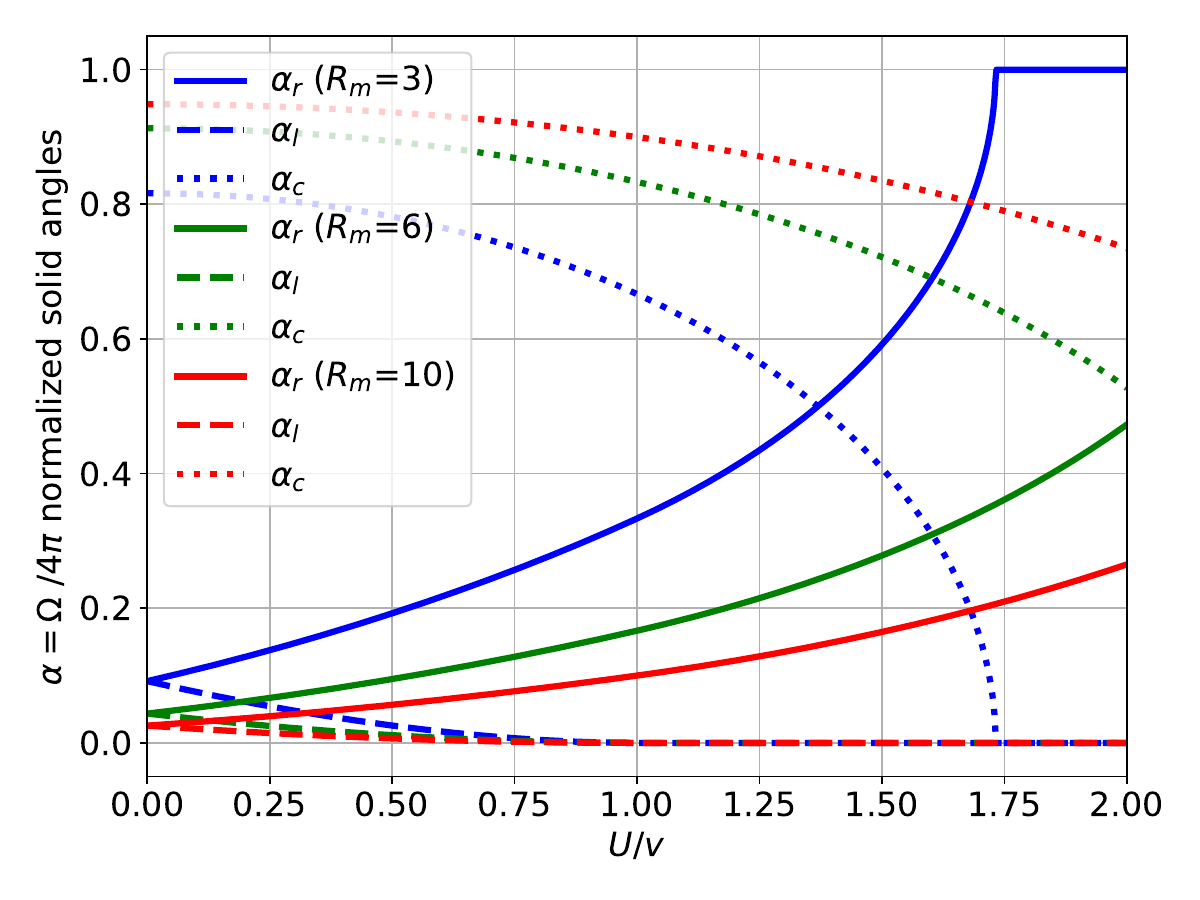}

    \caption{
    Analysis results for the effective loss cones in the laboratory frame.
    Left: the solutions of Eq.~(\ref{eq: v_z}) for the axial velocities, $v_{z,+}$ (solid black line) and $v_{z,-}$ (dashed line) on the loss cone boundaries as a function of the normalized MMM velocity $U/v$ for $R_m=3$. 
    The colored vertical lines indicate the boundaries between the different solution branches of Eqs.~(\ref{alpha_l})--(\ref{alpha_c}). 
    Center: the solutions of Eqs.~(\ref{alpha_l})--(\ref{alpha_c}) for the normalized solid angles of different populations, as a function of the normalized MMM velocity $U/v$ for $R_m=3$: right-going $\alpha_r$ (solid black line), left-going $\alpha_l$ (dashed line), and confined $\alpha_c$ (dotted).
    Right: same as the center panel but for three different values of the mirror ratio, $R_m=3,6,10$.
    }
    \label{fig: MMM loss cone solid angles}
\end{figure}

Here, $U_{m}=\sqrt{R_{m}-1}v$, 
denotes the velocity at which the transverse-velocity solution reaches its maximum.
For $U>U_{m}$, the modified right loss cone continues to expand as $U$ 
increases, eventually occupying the entire velocity sphere at $U=U_{last}$.
These expressions are plotted as a function of $U$ for several values of $R_m$ in the right panels of Fig.~\ref{fig: MMM loss cone solid angles}.

In Sec.~\ref{sec: rate equations} we will use the modified loss cone solid angles developed above, evaluated at $v=v_{th}$, which serves as the characteristic particle velocity. 
A more refined approach would average the solid angles over a Maxwellian velocity distribution.
Such an extension constitutes a higher-order correction and is therefore left to future work.

\section{Rate-equation model}
\label{sec: rate equations}

To quantify the confinement enhancement induced by adding MMM sections to a central magnetic mirror, we modify the semi-kinetic rate-equation model developed for stationary MM systems \cite{miller2021rate} to include the moving mirrors effects. 
For static MM systems, the rate-equation model splits the particles in each cell into three populations: confined particles and particles escaping through the right and left loss cones.
The corresponding population densities in the $i$'th cell are denoted $n_c^i$, $n_r^i$, and $n_l^i$.
Coulomb scattering drives intra-cell mixing between populations by redistributing particle velocity directions, while particles escaping through the loss cones undergo inter-cell transmission between neighboring cells.

The MMM rate-equations model retains the same physical processes as the stationary MM model, namely Coulomb scattering within each cell and inter-cell transmission through the loss cones. 
However, rather than modeling moving mirror cells directly, we treat the cells as stationary in the laboratory frame and represent the MMM dynamics through an additional effective transport term. 
This term captures the directed transport of confined particles between neighboring cells induced by the propagating mirror structure.
In this description, confined particles are propelled leftward toward the central cell, corresponding to transport from cell $i+1$ to cell $i$,
while the loss cone solid angle is modified to account for the transformation from the mirror frame to the lab frame, as derived in Sec.~\ref{sec: loss cone}.

The modified rate-equation model reads
\begin{eqnarray}
    \label{Eq: dn_c_dt} \dot n_{c}^i &=&  \nu_{s} \left[ \alpha_c (n_{l}^i + n_{r}^i) - \left( \alpha_r + \alpha_l \right) n_{c}^i\right] + \nu_\text{\tiny{MMM}} \left( n_{c}^{i+1} - n_{c}^i\right) \\
    \label{Eq: dn_tL_dt} \dot n_{l}^i &=&  \nu_{s}\left[\alpha_l (n_{r}^i+n_{c}^i) - \left( \alpha_r + \alpha_c \right) n_{l}^i\right] + \left( \nu_{t} + \nu_\text{\tiny{MMM}} \right) \left( n_{l}^{i+1} - n_{l}^i\right) \\                    
    \label{Eq: dn_tR_dt} \dot n_{r}^i &=& \nu_{s} \left[\alpha_r (n_{l}^i +n_{c}^i) - \left( \alpha_l + \alpha_c \right) n_{r}^i\right] + \nu_{t}  \left( n_{r}^{i-1} - n_{r}^i\right) + A\nu_\text{\tiny{MMM}} \left( n_{r}^{i+1} - n_{r}^i\right)  
\end{eqnarray}
where the effective solid angles of modified loss cones, $\alpha_{l,r,c}$, are defined in Eqs.~(\ref{alpha_l})--(\ref{alpha_c}) for the left-going, right-going, and confined populations, respectively. 
As in a static MM system, the transmission rate of loss cone particles between neighboring cells is given by $\nu_{t}=f_t v_{th}/l$, where $f_t=\left( T_i + T_e  \right)/T_i=2$ accounts for ambipolar transport under the assumption of ion-electron thermal equilibrium, $T_e=T_i$ and $v_{th}$ is the thermal velocity of the ions, representing the typical axial velocity of the right- and left-going particles in the lab frame, while the MMM drag effect is modeled by the coefficient $\nu_\text{\tiny{MMM}}$ (see below).

The ion-ion Coulomb scattering rate, $\nu_{s}$, depends on the thermodynamic conditions along the MM section \cite{miller2021rate}, and it is commonly treated under the isothermal assumption, which can be justified by rapid electron thermalization.
For fusion-relevant mirror plasmas, the corresponding mean free path (MFP), $\lambda=v_{th}/\nu_{s}$, is typically much larger than the cell length, \ie$\lambda\gg l$.
In this regime, particles escaping from the central cell would travel across the MM section essentially collisionlessly, without sufficient interactions to restore confinement.
Thus, an effective scattering mechanism is required to support the MMM pumping concept. 
Such mechanisms may be provided by externally applied RF-driven phase-space mixing fields, such as rotating electric \cite{miller2023rf} or rotating magnetic \cite{miller2026plugging} fields, or by relying on instability-driven anomalous scattering \cite{tolkachev2024electromagnetic}.  
For these processes, it is conceivable to expect an effective MFP of the order of a single MM cell, corresponding to one scattering event per cell traversal.
Accordingly, in the calculations below, we set $\lambda=l$, which corresponds to an enhancement of the effective scattering rate by approximately a factor of $2000$ relative to the typical Coulomb value.

\begin{figure}
    \centering
    \begin{minipage}{0.495\textwidth}
        \centering
        \includegraphics[clip, trim=0.0cm 0.0cm 0.0cm 0.50cm, width=\linewidth]{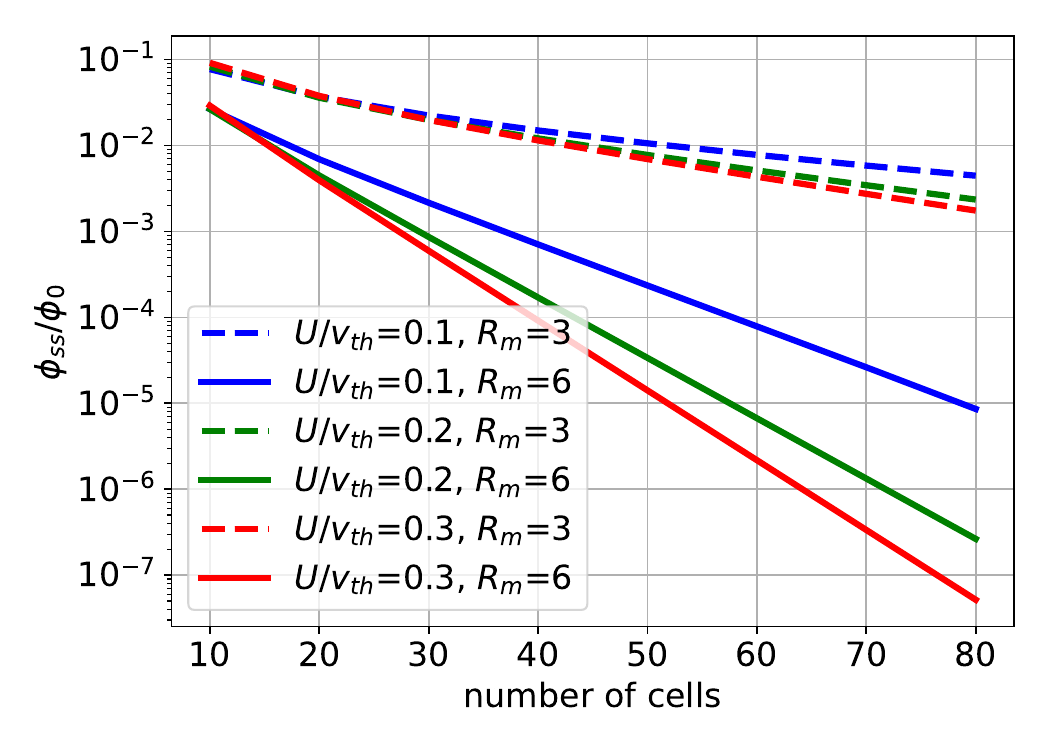}
    \end{minipage}
    \hfill
    \begin{minipage}{0.495\textwidth}
        \centering
        \includegraphics[clip, trim=0.0cm 0.0cm 0.0cm 0.50cm, width=\linewidth]{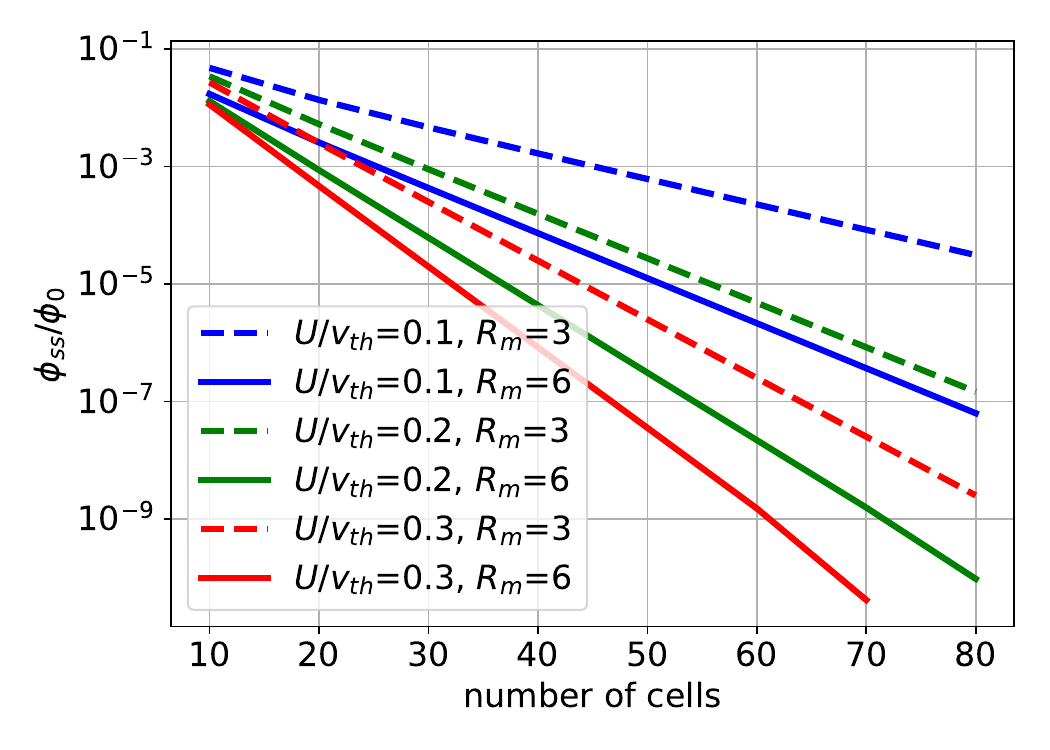}
    \end{minipage}
    \caption{Normalized steady state flux, $\phi_{ss}/\phi_{0}$ as a function of number of cells for several values of the mirror ratio, $R_m$, and mirror velocity, $U$. 
    The left and right panels correspond to the limiting cases of double-peaked ($A=0$) and single-peaked ($A=1$) velocity distributions, respectively.
    }
    \label{fig: MMM flux of N}
\end{figure}

The new components in the model are the last term in each of Eqs.~(\ref{Eq: dn_c_dt})-(\ref{Eq: dn_tR_dt}), representing the MMM-induced transport with an effective drag rate of $\nu_\text{\tiny{MMM}} = U/l$.
These terms arise from artificially modeling a moving MM system using the static MM rate-equation model.
Since the rate-equation model is formulated in the lab frame with static cells, it is natural to apply the MMM transport term to all three populations.
This is equivalent to assuming that the particle velocity distribution is Maxwellian, or more generally single-peaked, in the moving frame.
However, part of the right-going population may consist of particles originating from the central cell that have not yet scattered.
In this case, the overall velocity distribution will become double-peaked instead of Maxwellian.
The physical situation is expected to lie somewhere between these two extremes.
To account for both possibilities, we introduce a flag $A\in\{1,0\}$
corresponding to the limiting cases of a single-peaked (Maxwellian) and a double-peaked distribution, respectively, and study both limiting cases, which together provide a reasonable range for the physical solution.
As in \cite{miller2021rate} and \cite{miller2023rf}, we solve Eqs.~(\ref{Eq: dn_c_dt}-\ref{Eq: dn_tR_dt}), for the steady-state, \ie $\dot{n}_{c,l,r}=0$, and calculate the outgoing flux between two neighboring cells (\eg $i\rightarrow i+1$), which is given by

\begin{equation}
\phi_{i,i+1} \propto \nu_{t} \left( n_{r}^i - n_{l}^{i+1} \right)- \nu_\text{\tiny{MMM}} \left( n_{c}^{i+1} + n_{l}^{i+1} + An_{r}^{i+1} \right), 
\end{equation}

Fig~\ref{fig: MMM flux of N} presents the steady state flux as a function of the number of cells $N$ for several values of the mirror ratio $R_m$ and the MMM velocity $U$.
The fluxes shown in the figures are normalized by the single-mirror flux, $\phi_0 = n_0 v_{th}$, which provides a convenient reference.
Results are shown for the two limiting velocity distributions considered above, a double-peaked distribution ($A=0$, left panel) and a single-peaked distribution ($A=1$, right panel).
The figure shows that the steady-state flux decays exponentially with increasing number of cells, $N$. 
This behavior is robust and yields flux reductions of several orders of magnitude for reasonable parameter values. 
The corresponding decay rate, however, depends on $R_m$, $U$, and the choice of the parameter $A$.

\begin{figure}
    \centering
    \begin{minipage}{0.45\textwidth}
        \centering
        \includegraphics[clip, trim=0.0cm 0.5cm 0.0cm 0.5cm, width=\linewidth]{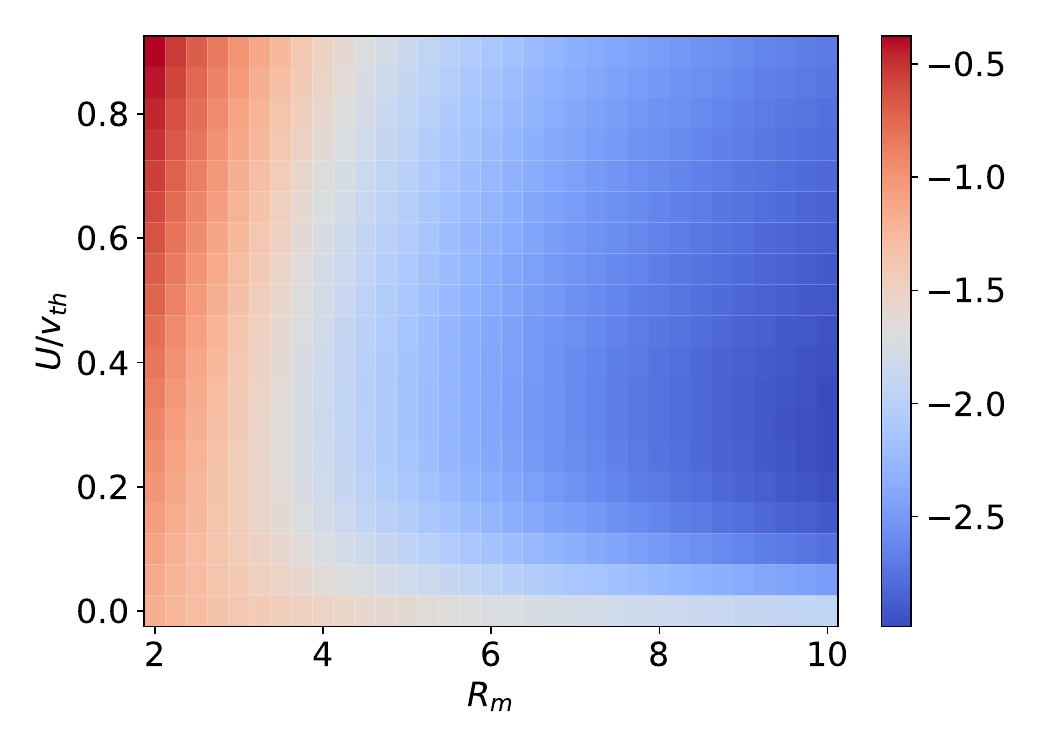}
    \end{minipage}
    \hfill
    \begin{minipage}{0.45\textwidth}
        \centering
        \includegraphics[clip, trim=0.0cm 0.5cm 0.0cm 0.5cm, width=\linewidth]{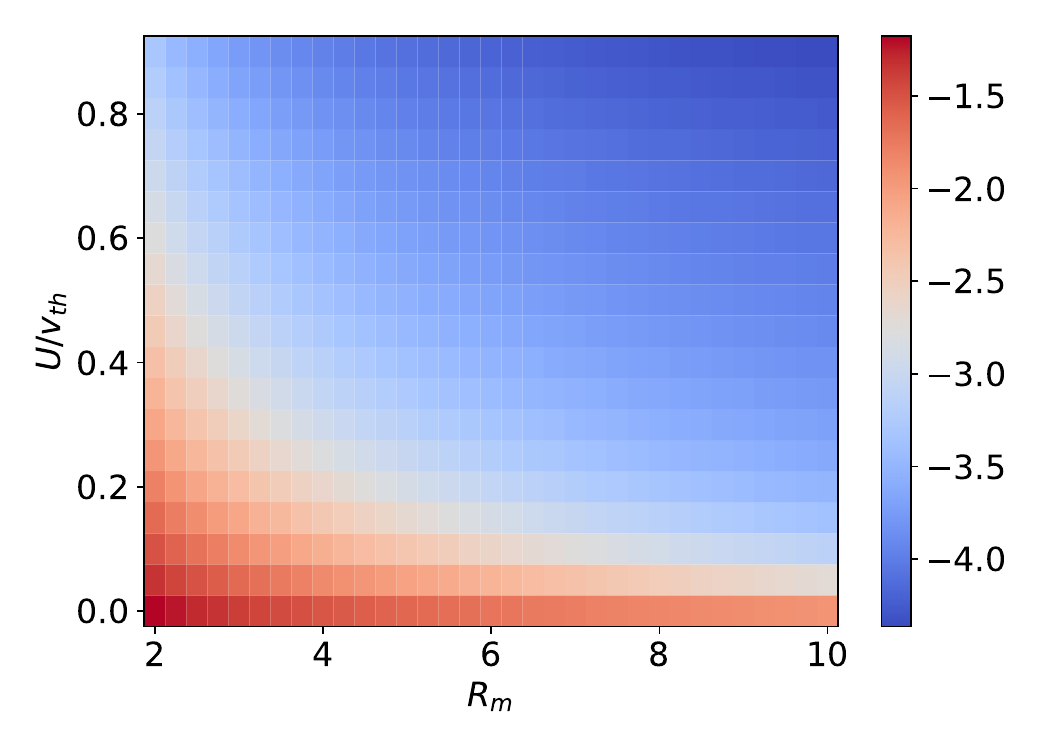}
    \end{minipage}
    \vspace{0.5cm}
    \begin{minipage}{0.45\textwidth}
        \centering
        \includegraphics[clip, trim=0.0cm 0.5cm 0.0cm 0.5cm, width=\linewidth]{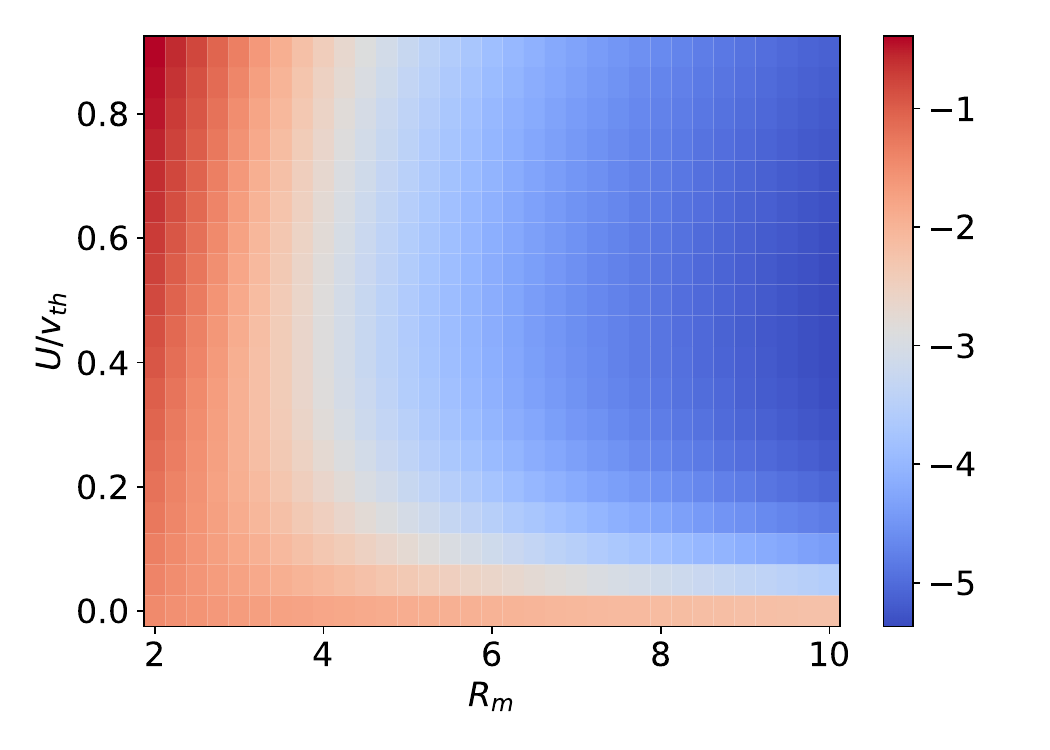}
    \end{minipage}
    \hfill
    \begin{minipage}{0.45\textwidth}
        \centering
        \includegraphics[clip, trim=0.0cm 0.5cm 0.0cm 0.5cm, width=\linewidth]{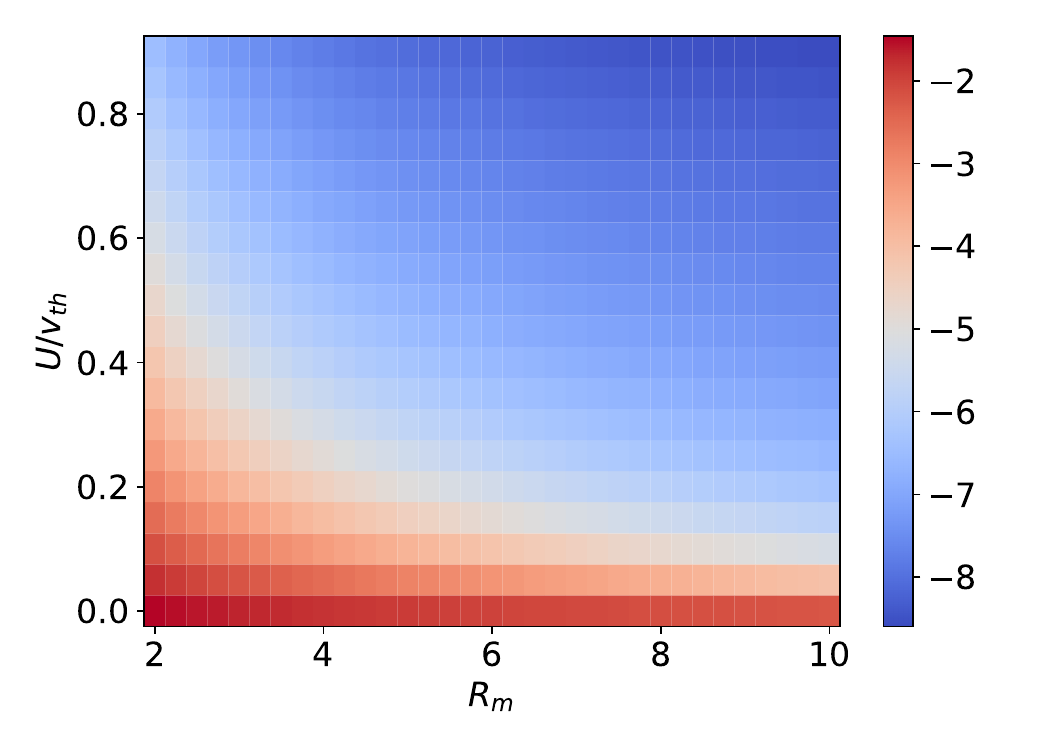}
    \end{minipage}
    \caption{Normalized steady state flux in the MMM, $\phi_{ss}/\phi_{0}$, as a function of $R_m$ and $U$.
    The top and bottom panels correspond to $N=20$ and $N=40$ cells, respectively. 
    The left and right panels show the two limiting velocity-distribution models, $A=0$ and $A=1$, respectively.
    The colormaps are logarithmic.}
    \label{fig: MMM flux of Rm,U}
\end{figure}

To further explore the MMM parameter space, we repeat the calculations and evaluate the steady-state flux as a function of $R_m$ and $U$ for fixed numbers of cells. 
The resulting flux maps are shown in Fig~\ref{fig: MMM flux of Rm,U} for $N=20$ (left panels) and $N=40$ (right panels).
The figures show that the assumed velocity-distribution model, represented by the parameter $A$, has a qualitative impact on the dependence of the steady state flux on $R_m$ and $U$ at fixed $N$. 
For $A=1$ (right panels), the steady-state flux exhibits a strictly monotonic decrease with increasing $U$ for all mirror ratios.
In contrast, for $A=0$ (left panels), a threshold-like transition is observed, where for mirror ratios below approximately $R_m \approx 4$, increasing $U$ surprisingly deteriorates confinement, as evidenced by the increase in the escaping flux.
This behavior can be understood as a consequence of the expansion of the right (escaping) loss cone with increasing $U$, as discussed in Sec.~\ref{sec: loss cone}.
In this regime, the enhanced escape flux outpaces the inward MMM transport, resulting in a net increase in the steady-state outgoing flux.
For sufficiently large mirror ratios, the larger confined population compensates for the growth of the loss cone, allowing the MMM transport to dominate and improve confinement.
This threshold effect is absent in the single-peaked case ($A=1$) because in this regime, the right-going (escaping) population also experiences the effective MMM transport.
Consequently, the increase in the escape flux due to the expanding loss cone is balanced by the corresponding increase in inward transport.
Thus, the confinement improves as $U$ increases for all values of $R_m$.

\section{Discussion and Conclusions}
\label{conclusions}

In this paper, we have introduced a microscopic analysis of the MMM concept, extending the fluid-model-based picture introduced in \cite{be2018plasma}. 
By constructing an explicit, Maxwell-consistent electromagnetic-field realization of an MMM system and following single-particle trajectories, we have directly demonstrated that trapped particles are swept by the MMM.
Analytically, we have derived expressions for the velocity-dependent loss-cone solid angles in the laboratory frame for a mirror moving at speed $U$. 
The resulting asymmetry between the upstream and downstream loss cones increases with $U$, thereby enlarging the escape channel in the direction opposite to mirror propagation. 
This intrinsic effect competes directly with MMM pumping, reducing the overall efficiency of flux suppression. 
However, when incorporating the MMM drag term together with the modified loss-cone fractions into a modified rate-equation model that was originally developed for a static multi-mirror system, we have found that the steady-state axial fluxes can be reduced by several orders of magnitude.
We further tested the sensitivity of the results to two limiting assumptions for the plasma velocity distribution: a single-peaked (drifting) distribution and a double-peaked distribution. 
In both cases, MMM pumping yields confinement improvements of several orders of magnitude, with the strongest effect observed in the drifting regime. 
This enhancement remains significant over wide ranges of mirror ratio, MMM velocity, and number of mirror cells.
For fusion-relevant parameters ($R_m\approx6$, $U\approx 0.1\, v_{\rm th}$, $N\gtrsim 30$), the end-loss flux is suppressed by $3$–$4$ orders of magnitude, exceeding the confinement improvement required to satisfy the Lawson criterion in mirror machines \cite{lawson1957some,miller2021rate}.
At the same time, the required voltages and currents are readily achievable with modern solid-state switch technology, the associated power cost remains modest on a reactor scale, and the predicted confinement gain is sufficiently large to justify near-term experimental demonstration.
Taken together, these results represent a significant step toward a firmer microscopic foundation for the MMM end-plug concept.

Finally, let us discuss the magnetic field configuration and its associated consequences and caveats.
In a naive design consisting just of MMM sections, particles initially confined in the central trap undergo repeated Fermi-like acceleration as the moving mirrors propagate inward, eventually driving them into the loss cones and degrading confinement.
To overcome this problem, we have added a static magnetic barrier separating the central (fusion) cell from the MMM sections.
The barrier restores conventional mirror motion in the central region and confines particles initially trapped in the central cell, preventing them from interacting with the moving fields.
Nonetheless, it was found that drifting particles bounce between the dynamic MMM and the static barrier fields, resulting in a squeezing effect that transiently increases their axial energy. 
As a result, and as discussed in Sec.~\ref{sec: single particle}, particles transported into the central cell are confined only by the dynamic MMM fields rather than by the static barrier.
Consequently, they remain susceptible to Fermi-like acceleration and eventual escape.
Sustained steady-state confinement, therefore, requires an additional scattering mechanism within the central cell to scatter incoming particles into the confined region of phase space associated with the static mirror barrier.

A second fundamental challenge, shared with conventional MM systems but exacerbated in MMM configurations, concerns particles that enter the loss cones. 
In the absence of sufficient scattering, such particles propagate through the MMM section essentially collisionlessly and are ultimately lost.
This problem becomes more severe in MMM systems because the effective outgoing loss cone increases with mirror velocity, as shown by the analytical loss-cone analysis. 
Consequently, particles are more likely to escape before being re-trapped, emphasizing the need for effective scattering mechanisms not only in the central cell but also throughout the MMM section.

These considerations highlight that the viability of MMM confinement ultimately depends on the presence of an effective scattering mechanism that continuously restores particles in the confined population in both the MMM sections and the central cell. 
In the absence of such a mechanism, particles transported into the central cell undergo progressive acceleration and eventual escape, while particles entering the loss cones propagate through the MMM sections without re-trapping and are ultimately lost from the system.
Therefore, for the fusion-relevant regime of weakly collisional plasmas with a large Coulomb MFP, identifying and quantifying possible effective scattering mechanisms, whether arising from turbulence-driven transport \cite{tolkachev2024electromagnetic} or externally applied RF fields \eg by a rotating magnetic field \cite{miller2026plugging}, remains a key challenge for future work.

Future work should also address the self-consistent plasma response, including ambipolar electric fields, micro-instabilities, collective drifts, and kinetic effects associated with finite-Larmor-radius and wave–particle interactions.
Despite the open questions discussed above, the present results demonstrate that MMM constitutes a promising and scalable route toward enhanced confinement in magnetic mirror fusion systems.

\data{
The data that support the findings of this study are available from the corresponding author upon reasonable request.}

\funding{This work was supported by the PAZI Foundation, Grant No. 2020-191, and by the Israeli Institute for Fusion Research, the Israeli Ministry of Energy, Grant No. R-25-4.} 


\bibliographystyle{iopart-num}
\bibliography{references}

\end{document}